\documentclass[reprint,amsmath,amsfonts,amssymb,aps,prx,preprintnumbers,superscriptaddress]{revtex4-1}

\usepackage[utf8]{inputenc}
\usepackage[T1]{fontenc}

\usepackage{graphicx}
\usepackage{tabularx}
\usepackage{multirow}
\usepackage{dcolumn}
\usepackage{booktabs}
\usepackage{newtxtext}
\usepackage[cmbraces]{newtxmath}
\usepackage{bm}

\usepackage{color}
\definecolor{darkblue}{rgb}{0,0,.6}

\definecolor{greencomment}{rgb}{0,.6,0}

\usepackage{ifpdf}
\ifpdf
  \usepackage{epstopdf}
  \usepackage[pdftex,unicode,pdfstartview={FitH},pdfborder={0 0 0}]{hyperref}
  \usepackage{hypcap}
\else
  \usepackage[hypertex]{hyperref}
\fi
\hypersetup{
  bookmarksnumbered = true,
  colorlinks = true, linkcolor = darkblue,
  citecolor = darkblue, filecolor = darkblue,
  menucolor = darkblue, urlcolor = darkblue
}

\newcolumntype{C}{>{$\displaystyle}c<{$}}
\newcolumntype{R}{>{$\displaystyle}r<{$}}

\newcommand{\bra}[1]{\ensuremath{\left\langle#1\right|}}
\newcommand{\ket}[1]{\ensuremath{\left|#1\right\rangle}}

\let\vec=\mathbf

\begin{document}

\title{Exciton g-factors in monolayer and bilayer WSe$_2$ from experiment and theory}

\author{Jonathan F\"orste}
\def\LMU{Fakult\"at f\"ur Physik, Munich Quantum Center, and Center for NanoScience (CeNS), Ludwig-Maximilians-Universit\"at M\"unchen, Geschwister-Scholl-Platz 1, D-80539 M\"unchen, Germany}
\affiliation{\LMU}

\author{Nikita~V.~Tepliakov}
\affiliation{Chair of Computational Condensed Matter Physics (C3MP), Institute of Physics, Ecole Polytechnique F\'ed\'erale de Lausanne (EPFL), CH-1015 Lausanne, Switzerland}
\affiliation{Information Optical Technologies Center, ITMO University, 197101 Saint Petersburg, Russia}

\author{Stanislav~Yu.~Kruchinin}
\affiliation{Center for Computational Materials Sciences, Faculty of Physics, University of Vienna, Sensengasse 8/12, A-1090 Vienna, Austria}

\author{Jessica Lindlau}
\author{Victor Funk}
\author{Michael F\"org}
\affiliation{\LMU}

\author{Kenji Watanabe}
\author{Takashi Taniguchi}
\affiliation{National Institute for Materials Science, Tsukuba, Ibaraki 305-0044, Japan}

\author{Anvar~S.~Baimuratov}
\affiliation{\LMU}

\author{Alexander H\"ogele}
\affiliation{\LMU}
\affiliation{Munich Center for Quantum Science and Technology (MCQST),
Schellingtr. 4, D-80799 M\"unchen, Germany}

\date{\today}

\begin{abstract}
The optical properties of monolayer and bilayer transition metal dichalcogenide semiconductors are governed by excitons in different spin and valley configurations, providing versatile aspects for van der Waals heterostructures and devices. We present experimental and theoretical studies of exciton energy splittings in external magnetic field in neutral and charged WSe$_2$ monolayer and bilayer crystals embedded in a field effect device for active doping control. We develop theoretical methods to calculate the exciton $g$-factors from first principles and tight-binding for all possible spin-valley configurations of excitons in monolayer and bilayer WSe$_2$ including valley-indirect exciton configurations. Our theoretical and experimental findings shed light on some of the characteristic photoluminescence peaks observed for monolayer and bilayer WSe$_2$. In more general terms, the theoretical aspects of our work provide new guidelines for the characterization of single and few-layer transition metal dichalcogenides, as well as their heterostructures, in the presence of external magnetic fields.
\end{abstract}


\maketitle

\section{Introduction}

Monolayer (ML) and bilayer (BL) transition metal dichalcogenides (TMDs) such as WSe$_2$ represent semiconductor building blocks for novel van der Waals heterostructures. By virtue of sizable light-matter coupling governed by excitons \cite{wang_colloquium_2018} they exhibit versatile potential for applications in photonics and optoelectronics \cite{Wang2012,mak_photonics_2016}, opto-valleytronics \cite{Neumann2017,Schaibley2016} and polaritonics \cite{schneider_two-dimensional_2018}. Most recently, the optical interface to TMDs has been instrumental for the observation of strongly correlated electron phenomena in twisted homo- and heterobilayer moir\'{e} systems \cite{tang_wse2ws2_2019,regan_optical_2019,shimazaki_moire_2019}. 

The key to further developments of van der Waals heterostructures for fundamental studies and practical devices using TMD MLs and BLs is the detailed understanding of their optical properties. While substantial understanding of zero-momentum excitons in ML and BL WSe$_2$ has been established \cite{wang_colloquium_2018}, some important aspects remain subject of debate \cite{Koperski2017}. This holds, in particular, for valley-dark excitons with finite center-of-mass momentum that escape direct optical probes by virtue of momentum mismatch with photons. In MLs, they complement the notion of intravalley spin-bright and spin-dark excitons \cite{wang_colloquium_2018}, and they entirely dominate the photoluminescence (PL) from the lowest-energy states in native homobilayers of WSe$_2$ \cite{Lindlau2017BL}.  

Within the realm of optical spectroscopy techniques, magneto-spectroscopy provides means for studying the exciton spin and valley degrees of freedom. Magneto-luminescence experiments on ML WSe$_2$ in the presence of out-of-plane and in-plane magnetic fields, for instance, have been used to quantify the valley Zeeman splitting of bright excitons \cite{Srivastava2015,Aivazian2015,Wang2015,Mitioglu2015,Stier2018,Koperski2018}, or to brighten spin-dark excitons \cite{Zhang2017,Molas2017,RobertPRB2017}, respectively. 
To date, however, a rigorous assignment of exciton $g$-factors to intervalley excitons with finite momentum falls short mainly due to the lack of theoretical predictions \cite{Koperski2017}.

In this work, we develop theoretical methods to evaluate $g$-factors for excitons in different spin and valley configurations, and provide explicit values for WSe$_2$ ML and BL excitons composed from electron and hole states away from high symmetry points of the first Brillouin zone. Our calculations go beyond the existing tight-binding (TB) models by employing the density functional theory (DFT). We compare our theoretical results with experimentally determined $g$-factors of intravalley excitons, and use them to interpret ambiguous peaks in the PL spectra of ML and BL WSe$_2$ attributed to intervalley excitons. 

\section{Experimental Results}

\begin{figure}[t]
\centering
\includegraphics[scale=0.9]{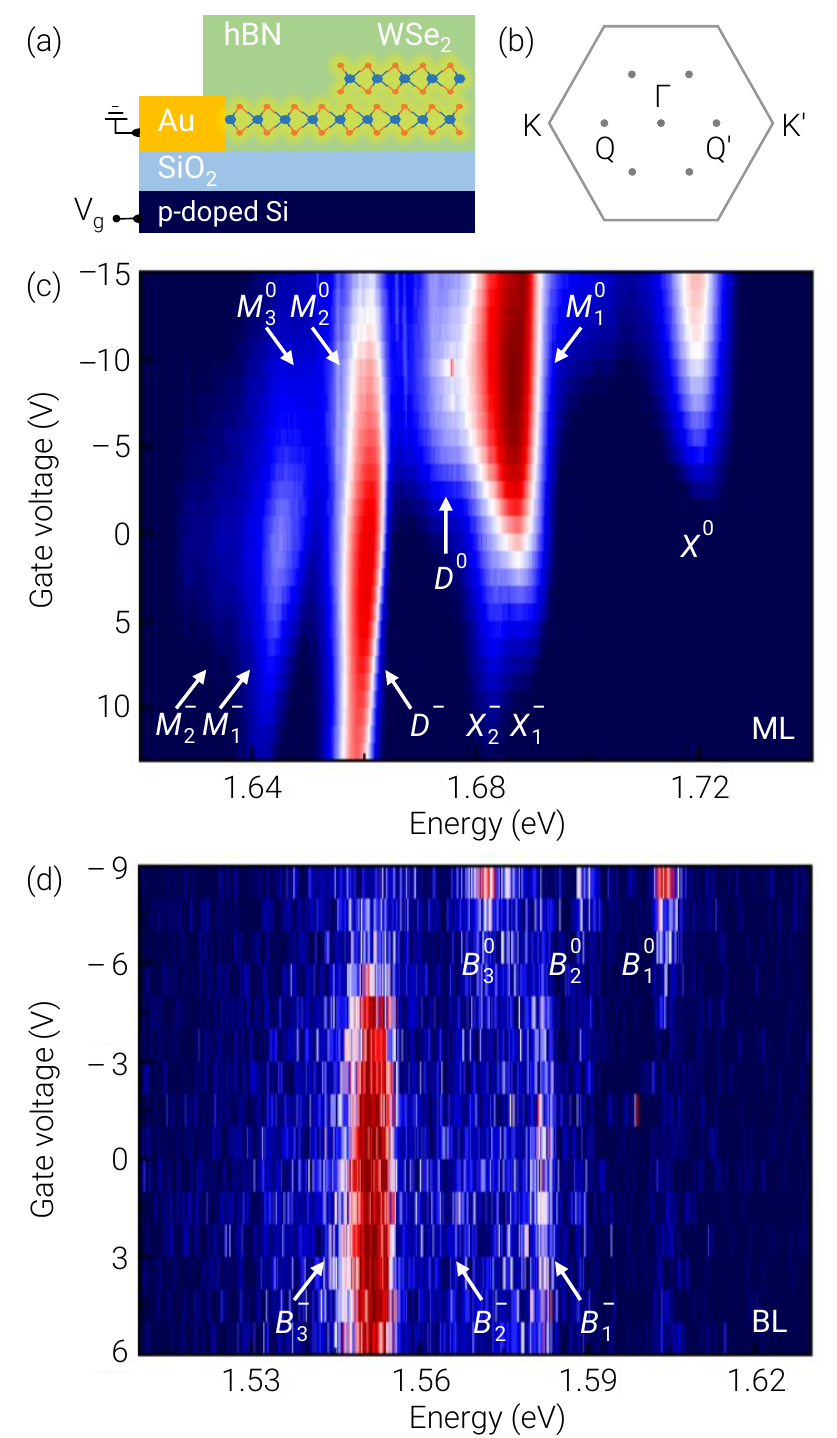}
\caption{Logarithmic false-color plots of the photoluminescence as a function of the gate voltage, recorded at representative positions of (a) monolayer and (b) bilayer WSe$_2$ regions of the same field-effect device under laser excitation at $1.85$~eV. The upper and lower dashed lines indicate the regimes of charge neutrality and negative doping, respectively, with the corresponding magneto-luminescence spectra shown in Fig.~2. The monolayer exhibits characteristic photoluminescence peaks of neutral bright ($X^0$) and dark ($D^{0}$) excitons as well as the negatively charged bright trion doublet ($X^{-}_1$ and $X^{-}_2$) and dark trion ($D^{-}$) emission peaks. All other peaks of monolayer ($M$) and bilayer ($B$) photoluminescence are labelled according to their charge state in the superscript and an increasing subscript number.}
\label{fig1}
\end{figure}

In our experiments we used a field-effect heterostructure based on an exfoliated WSe$_2$ crystal with extended ML and  BL regions that was encapsulated in hexagonal boron nitride (hBN). The device layout is shown schematically in Fig.~\ref{fig1}(a), and the first Brillouin zone of ML and BL WSe$_2$ with most relevant points in Fig.~\ref{fig1}(b). To control the charge doping level, the crystal was contacted by a gold electrode deposited on a $50$~nm thick thermal silicon oxide layer of a highly p-doped silicon substrate. With the electrode grounded, a voltage applied to the doped silicon back gate was used to control the doping in both ML and BL regions. The sample was mounted in a cryogenic confocal microscope and cooled down in a closed-cycle magneto-cryostat with a base temperature of $3.2$~K. The PL was excited with a continuous-wave laser diode at $1.85$~eV focussed to the diffraction-limited confocal excitation and detection spots of a low-temperature apochromatic objective, dispersed with a monochromator and detected with a nitrogen-cooled CCD. The excitation power of a few $\mu$W was kept below the regimes of neutral and charged biexcitons \cite{you_observation_2015,barbone_charge-tuneable_2018,steinhoff_biexciton_2018,li_revealing_2018}. Magneto-luminescence experiments were performed in Faraday configuration with a bi-directional solenoid at magnetic fields of up to $9$~T.

The evolution of the PL with the gate voltage is shown in Fig.~\ref{fig1}(c) and (d) for representative spots of ML and BL regions, respectively. In Fig.~\ref{fig1}(c), the ML reaches the intrinsic limit at gate voltages below $-5$~V consistent with residual n-doping of the exfoliated crystal \cite{Mak2013,Courtade-PRB2017}. The neutral regime is characterized by the bright exciton PL ($X^{0}$) at $1.72$~eV and a series of red-shifted peaks that we label as $M^0_1$, $M^0_2$ and $M^0_3$. None of these peaks with respective red-shifts of $35$, $60$ and $75$~meV from the bright exciton peak is to be attributed to the PL of dark excitons ($D^{0}$) with $42$~meV red-shift \cite{Zhou2017,Wang2017,RobertPRB2017}. In our sample, this feature is a rather weak shoulder at the low-energy side of $M^0_1$. At positive gate voltages, the ML is charged with electrons and thus exhibits the characteristic signatures of a bright trion doublet ($X^{-}_1$ and $X^{-}_2$) split by the exchange energy of $\sim 6$~meV \cite{Courtade-PRB2017}, the dark trion ($D^{-}$) at $28$~meV red-shift from $X^{-}_1$ \cite{LiuPRL2019,LiNanoLett2019,LiuPRR2019,HeNatComm2020}, and a series of low-energy peaks dominated by the peak $M^{-}_1$ at $44$~meV red-shift \cite{LiuPRR2019,HeNatComm2020}. 

\begin{figure}[t]
\centering
\includegraphics[scale=0.9]{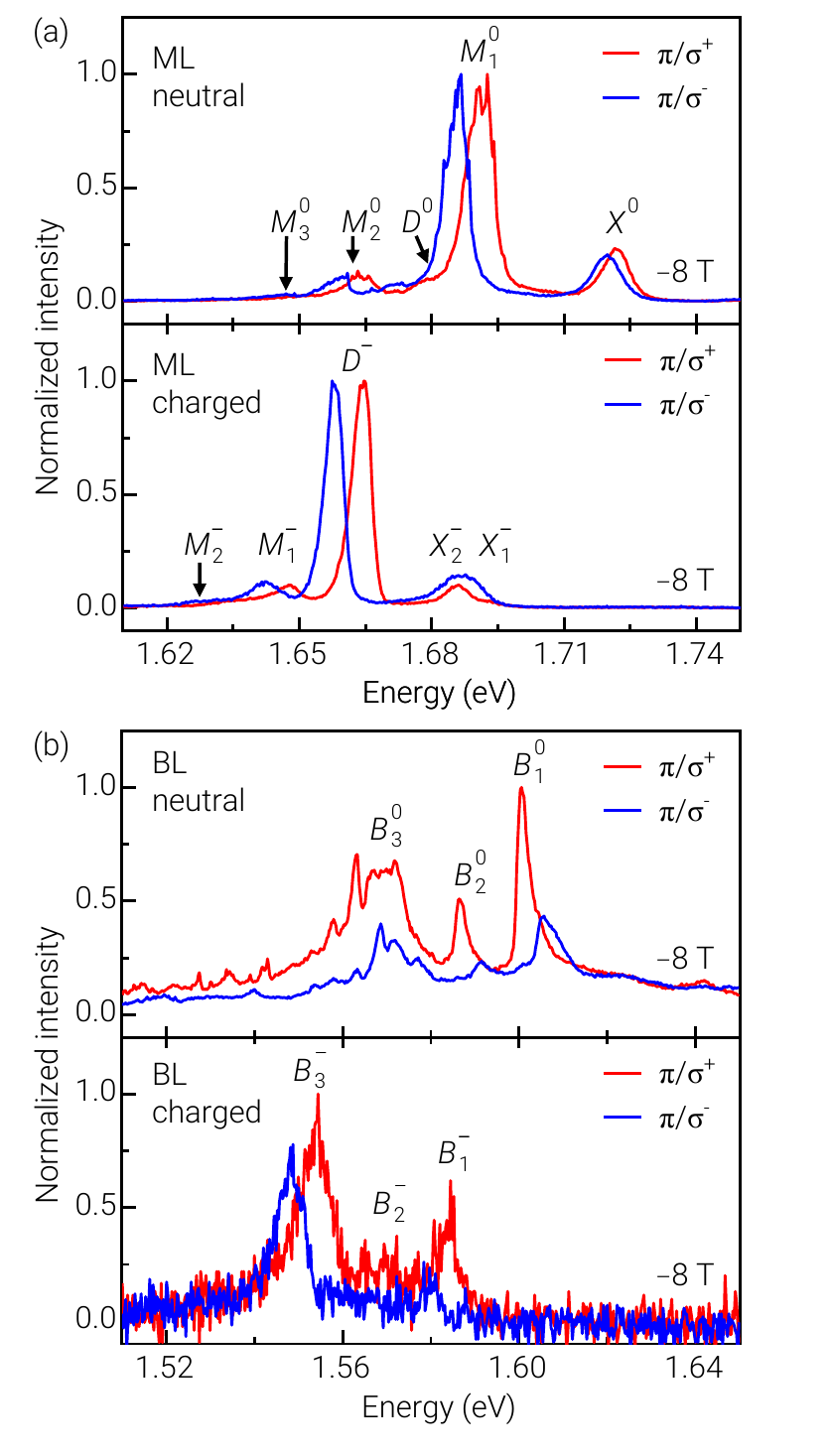}
\caption{Photoluminescence spectra of (a)~monolayer and (b)~bilayer WSe$_2$ in a perpendicular magnetic field of $-8$~T. The neutral and negatively charged regimes are shown in the top and bottom panels, respectively. The spectra were recorded with linearly polarized excitation ($\pi$) and circularly polarized detection ($\sigma^+$ and $\sigma^-$).}
\label{fig2}
\end{figure}

\begin{figure}[t!]
\centering
\includegraphics[scale=0.95]{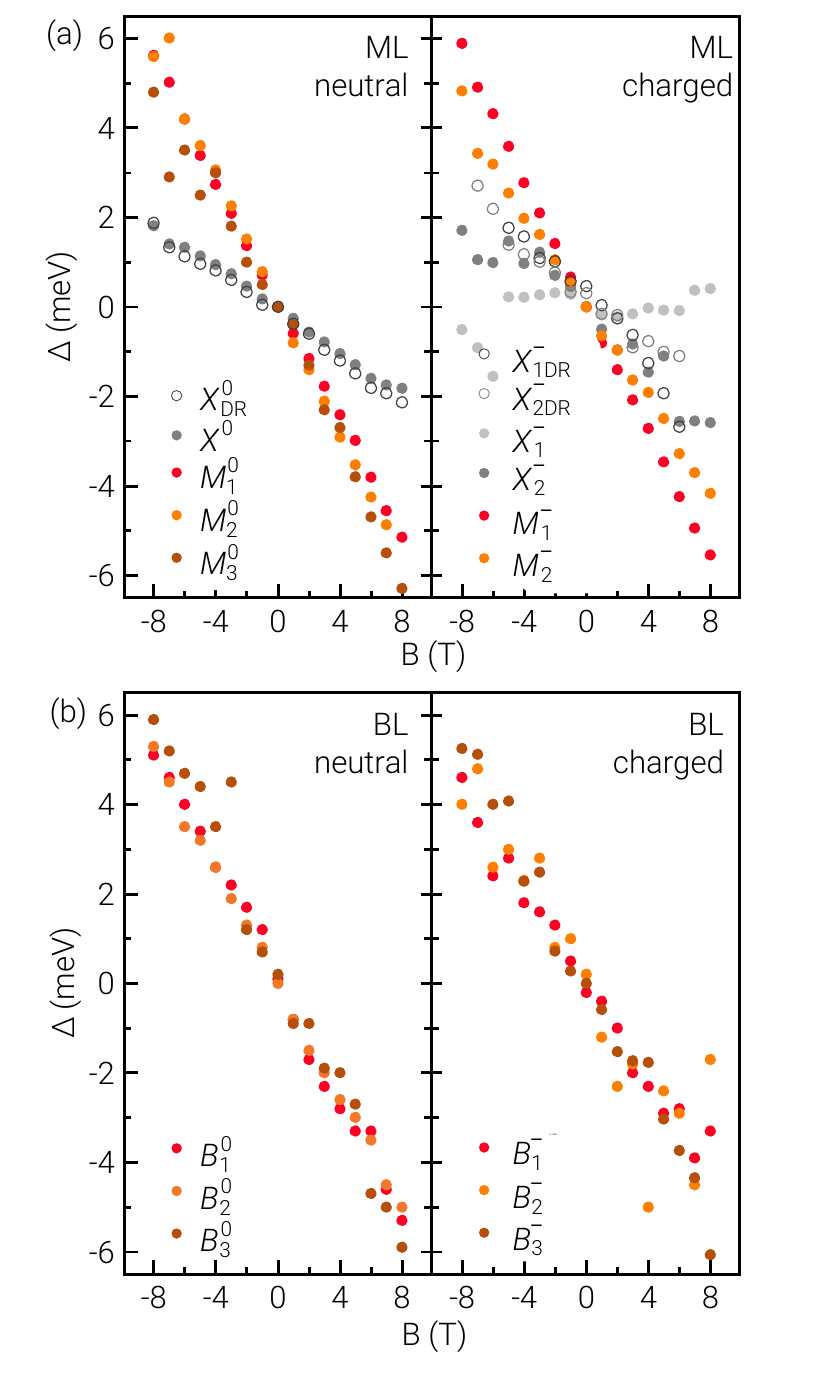}
\caption{Valley Zeeman splitting $\Delta E$ as a function of the magnetic field for the photoluminescence peaks (closed circles) of (a) monolayer and (b) bilayer WSe$_2$ in the neutral (left panel) and negatively charged (right panel) regimes, respectively. Complementary data (open circles) are from polarization-resolved reflectivity.}
\label{fig3}
\end{figure}

The PL from the BL region in Fig.~\ref{fig1}(d) is characterized by a multi-peak structure, more than $100$~meV below $X^{0}$. It exhibits the same limits of charge neutrality and electron doping as a function of the gate voltage, consistent with the charging behavior of the ML in Fig.~\ref{fig1}(c). The BL peaks, labelled by an increasing subscript number with decreasing peak energy as $B^{0}_1$ through $B^{0}_3$ and $B^{-}_1$ through $B^{-}_3$ in the neutral and negative regime, respectively, correspond to phonon-sidebands of neutral and charged momentum-indirect excitons with a global red-shift of $22$~meV at about $-7$~V in Fig.~\ref{fig1}(d) \cite{Lindlau2017BL}. According to the single-particle band structure of BL WSe$_2$ \cite{Wickramaratne2014,Terrones2014}, the field-induced electron concentration is accommodated at the conduction band edge by the six inequivalent $Q$-valleys. However, the nature of the hole states that constitute the lowest-energy momentum-dark excitons as long-lived reservoirs of phonon-assisted PL remains ambiguous. The energetic proximity of the valance band edge states at $K$ and $\Gamma$ in BL WSe$_2$ \cite{Wilson2017} renders $QK$ and $Q\Gamma$ excitons and trions (composed from electrons at $Q$ and holes at $K$ or $\Gamma$) nearly degenerate, which in turn complicates their energetic ordering \cite{Lindlau2017BL}. 

To examine the origin of the BL peaks, and to shed light on the nature of ML peaks with ambiguous or partly controversial interpretation, we performed magneto-spectroscopy in the two well-defined limits of charge neutrality and negative doping. The external magnetic field $B$ was applied along the $z$-axis perpendicular to the sample. It removes the valley degeneracy and splits the exciton reservoirs by their characteristic Zeeman energies proportional to the exciton $g$-factor in WSe$_2$ \cite{Srivastava2015,Aivazian2015,Wang2015,Mitioglu2015,Stier2018,Koperski2018}. The respective polarization-contrasting spectra recorded at $-8$~T under linearly polarized excitation ($\pi$) and circularly polarized detection ($\sigma^+$ and $\sigma^-$) for the neutral (negatively charged) ML and BL are shown in the top (bottom) panel of Fig.~\ref{fig2}(a) and (b). 

At each magnetic field, we quantified the experimental Zeeman splitting for every PL peak as the energy difference $\Delta =E^{+}-E^{-}$ between the peak energies $E^{+}$ and $E^{-}$ recorded under $\sigma^+$ and $\sigma^-$ polarized detection. The left and right panels of Fig.~\ref{fig3}(a) and (b) show $\Delta$ as a function of the magnetic field for all peaks of the neutral and negatively charged ML and BL, respectively. The set of data derived from magneto-PL measurements was complemented for $X^0$, $X^-_1$, and $X^-_2$ by performing magneto-reflectivity under circular excitation and detection. The corresponding experimental exciton $g$-factors, obtained from $\Delta=g \mu_{B}B$ as the slopes of best linear fits to the data in Fig.~\ref{fig3} scaled by the Bohr magneton $\mu_B$, are summarized in Tab.~\ref{tab1}. The negative sign of the $g$-factors reflects the energy ordering of exciton states that exhibit higher (lower) energy for $\sigma^-$ ($\sigma^+$) polarized PL peaks at positive magnetic fields.

\begin{table}
\caption{Experimental $g$-factors obtained from magneto-luminescence (\footnote{$X^0: -4.3 \pm 0.1$; $X^-_1: -4.7 \pm 0.3$; $X^-_2: -6.5 \pm 0.4$}complementary data from magneto-reflectivity) of neutral and negatively charged monolayer and bilayer WSe$_2$.}
  \centering    
  \begin{ruledtabular}
  \begin{tabular}{RRRRR|RRR}
       \multicolumn{4}{c}{ML}             && \multicolumn{3}{c}{BL} \\
\hline
        X^0   & M^0_1    & M^0_2    & M^0_3    && B^0_1    & B^0_2    & B^0_3  \\
        -4.1   & -11.5   & -12.6   & -11.4   && -11.4   & -10.8   & -12.8   \\
    \pm 0.1 & \pm 0.1 & \pm 0.2 & \pm 0.4 && \pm 0.2 & \pm 0.1 & \pm 0.2 \\   
\hline
      X^-_1   &X^-_2    & D^-    & M^-_1    && B^-_1    & B^-_2    & B^-_3    \\
      -4.6   & -1.3    & -12.2   & -9.0    && -9.1    & -9.8    & -11.5  \\
   \pm 0.3 & \pm 0.3 & \pm 0.1 & \pm 0.1 && \pm 0.3 & \pm 1.0 & \pm 0.4 \\
  \end{tabular}
  \end{ruledtabular}
\label{tab1}
\end{table}

In ML WSe$_2$, the $g$-factors of both neutral and negatively charged excitons with the corresponding PL peaks $X^0$, $D^0$, $X^-_1$, $X^-_2$ and $D^-$ have been established in previous experiments on a wide range of different samples \cite{Srivastava2015,Aivazian2015,Wang2015,Mitioglu2015,Zhang2017,Molas2017,RobertPRB2017,Stier2018,Koperski2018,LiuPRL2019,LiNanoLett2019,LiuPRR2019,HeNatComm2020}. Our results for the bright exciton and the trion doublet in Tab.~\ref{tab1} agree well with these reports if we discard the magneto-luminescence result for $X^{-}_{1}$ that is compromised by both a vanishingly small PL intensity at high magnetic fields and the relatively broad linewidth of $6$~meV in our sample. Due to this inhomogeneous broadening we are unable to track the dispersion of the relatively weak spin-dark exciton peak $D^0$, with $g$-factors ranging between $9.1$ and $9.9$ in previous reports \cite{RobertPRB2017,LiuPRL2019,LiuPRR2019,HeNatComm2020} nor its chiral-phonon replicum with the same $g$-factor at $65$~meV red-shift from $X^0$ \cite{LiNatCommun2019,LiuPRR2019,HeNatComm2020}. The signature of the latter is overwhelmed in our spectra by the peak $M^0_2$ with $60$~meV red-shift and a $g$-factor of $-12.9 \pm 0.7$ in agreement with values reported from samples with spectrally narrow PL \cite{LiuPRR2019,HeNatComm2020}. The red-most peak $M^0_3$ features the same $g$-factor within the experimental error bars as $M^0_1$, suggesting a joint reservoir as their origin. The negatively charged trion $D^-$ was reported to have the same $g$-factor as its neutral counterpart \cite{LiuPRL2019,LiNanoLett2019,LiuPRR2019,HeNatComm2020}, whereas we determine $-12.2\pm0.1$. The agreement with previous reports is better for the peak $M^-_1$ with a $g$-factor of $-9.0\pm 0.1$ that is supposed to be a phonon sideband of $D^-$ \cite{LiuPRR2019,HeNatComm2020}. The latter studies also reported an intense PL peak between $M^-_1$ and  $D^-$ with a remarkably small $g$-factor of $-4.1$ \cite{LiuPRR2019} and $-3.4$ \cite{HeNatComm2020}. This peak of unidentified origin is missing in our spectra from the negative doping regime. 

There are other peaks in ML WSe$_2$ without conclusive assignment, and in particular $M^0_1$ has received controversial interpretation as phonon-assisted PL from virtual trions \cite{VanTuan2019}, phonon sidebands of momentum-dark $Q$-excitons \cite{Lindlau2017ML}, or zero-phonon PL of finite-momentum excitons in spin-like configuration \cite{HeNatComm2020} that we denote as $K'_L$. Due to the lack of theory for the $g$-factors of excitons with finite center-of-mass momentum, the task of confronting the competing hypotheses with the characteristic valley Zeeman splittings of controversial ML peaks has remained elusive. The same shortcoming holds for both neutral and charged BL excitons with finite center-of-mass momentum. To shed additional light on the nature of PL peaks in both ML and BL WSe$_2$, we calculate in the following the $g$-factors for excitons in different spin and valley configurations from TB and DFT. 

\section{THEORETICAL FORMALISM}

  \begin{table*}[htb!]
\caption{Two sets of tight-binding fitting parameters for monolayer WSe$_2$: TB-1 from Ref.~\onlinecite{Fang2015} and TB-2 from Ref.~\onlinecite{Rybkovskiy2017} (in units of eV).}
  \begin{ruledtabular}
  \begin{tabular}{CRRRRRRRRRRRRR}
   \textrm{Set}& t_{6,6}^{(1)} & t_{7,7}^{(1)}  & t_{8,8}^{(1)}  & t_{6,7}^{(1)} & t_{6,8}^{(1)}  & t_{7,8}^{(1)} & t_{9,9}^{(1)}  & t_{10,10}^{(1)} & t_{11,11}^{(1)} & t_{9,10}^{(1)} & t_{9,11}^{(1)} & t_{10,11}^{(1)} &                  \\
                  & t_{9,6}^{(5)} & t_{11,6}^{(5)} & t_{10,7}^{(5)} & t_{9,8}^{(5)} & t_{11,8}^{(5)} & t_{9,6}^{(6)} & t_{11,6}^{(6)} & t_{9,8}^{(6)}   & t_{11,8}^{(6)}  & \epsilon_6     & \epsilon_{7,8} & \epsilon_{9}    & \epsilon_{10,11} \\
    \hline
    \textrm{TB-1} & -0.3330       & 0.3190         & -0.5837        & -0.1250       & 0.4233         & -0.2456       & -0.2399        & 1.0470          & 0.0029          & 0.1857         & -0.0377        & -0.1027         &                  \\
                  & -0.8998       & -0.9044        & 1.4030         & -0.8548       & 0.5711         & -0.0676       & -0.1608        & -0.2618         & -0.2424         & -0.1667        & 0.0984         & -3.3642         & -2.1820          \\
                  \addlinespace
    \textrm{TB-2} & -0.985        & 0.618          & -0.775         & -0.853        & -0.0083     & -0.412        & 0.191          & 1.22            & 0.028           & -0.176         & -0.039         & -0.228          &                  \\
                  & -0.8          & 0.183          & 1.8            & 0.811         & -0.0766        & 0.993         & 0.728          & 0.888           & 0.31            & -0.999         & -0.047         & -1.69           & -2.25            \\
\end{tabular}
\end{ruledtabular}
\label{tab3}
\end{table*}

We consider a crystal electron in a Bloch state $\psi_{n\mathbf k} (\mathbf r) = S^{-1/2} \exp(i\mathbf{kr}) u_{n\mathbf k} (\mathbf r)$ with energy $E_{n \mathbf k}$, where $n$ is the band number, $\mathbf k$ is the wave vector, $u_{n\mathbf k} (\mathbf r)$ is the periodic Bloch amplitude, and $S$ is the normalization area. In the presence of a weak perturbation by a static magnetic field $\mathbf B$, the first-order correction to the electron energy is proportional to $\mathbf B$ and given by \cite{Landau2013}:
\begin{equation}
  V_n (\mathbf k) =
    \mu_\textrm{B} \mathbf{B}
    [g_0 \mathbf{s}+\mathbf{L}_n (\mathbf k)],
    \label{eq_Landau}
\end{equation}
where $\mu_\textrm{B}  = |e|\hbar/(2 m_0 c)$ is the Bohr magneton, $e$ and $m_0$ are the charge and mass of the free electron, and $c$ is the speed of light. The expression in square brackets is usually called the effective magnetic moment \cite{Roth1959,Bir1974}, which contains both spin and orbital contributions. In particular the first term is proportional to the free electron Land\'e factor $g_0 \simeq 2$ and the spin angular momentum $\mathbf s = \bm\sigma/2$, where $\bm\sigma$ denotes the Pauli matrix. 

The second term, $\mathbf{L}_n (\mathbf k) = \bra{\psi_{n\mathbf k} (\mathbf r)} \mathbf L \ket{\psi_{m\mathbf k} (\mathbf r)}$, is the orbital angular momentum with the operator $\mathbf L = \hbar^{-1} [\mathbf{r} \times \mathbf p]$. To obtain its matrix elements one can reduce the calculation to the interband matrix elements of the momentum operator $\mathbf{p}$ or the operator of the space coordinate $\mathbf{r}$  \cite{Roth1959,Bir1974,Xiao2010,Wang2015}:
\begin{align}
\label{eq-veltblz}
  \mathbf{L}_n (\mathbf k) &=
    \frac{1}{i m_0}
    \sum_{m \neq n}
      \frac{\left[\mathbf p_{nm} (\mathbf k) \times \mathbf p_{mn} (\mathbf k)\right]}{E_{n \mathbf k} - E_{m \mathbf k}},  
\\
\label{eq_vecl}
\mathbf{L}_n (\mathbf k) &=
    \frac{m_0}{i \hbar^2}
    \sum_{m \neq n}
      [\bm{\xi}_{nm} (\mathbf k) \times \bm{\xi}_{mn}(\mathbf k)]
      (E_{n \mathbf k} - E_{m \mathbf k}),
\end{align}
where $m$ is the sum over all bands but the band of interest, and $\bm{\xi}_{nm} (\mathbf k) = i \bra{u_{n\mathbf k} (\mathbf r)} \partial / \partial \mathbf k \ket{u_{m\mathbf k} (\mathbf r)}$ is the interband matrix element of the crystal coordinate operator.

In the following, we restrict our analysis to the orientation of the magnetic field along the $z$-axis and define the electron Zeeman splitting as the difference between the energy of the electron state with wave vector $+\mathbf{k}$ and spin projection $+s$ along the $z$-axis and the state with $-\mathbf{k}$ and $-s$ as:
\begin{equation}
  \Delta_n (\mathbf k) =
    V_n (+\mathbf k) - V_n (-\mathbf k) =
      2 \mu_\textrm{B} B [g_0 s +L_n (\mathbf k)].
\end{equation}
Thus, the electron $g$-factor of Bloch electrons in the $n$-th band can be written as:
\begin{equation}
  g_n (\mathbf k) =
    \frac{\Delta_n (\mathbf k)}{\mu_\textrm{B} B} =
      \pm g_0 + 2 L_n (\mathbf k)
\end{equation}
with $+$ ($-$) for $s=+1/2$ ($-1/2$) corresponding to spin up~(down) projections along $z$ denoted as $\uparrow$~($\downarrow$), and the explicit expression for the $z$-component of the orbital angular momentum:
\begin{align} 
\label{eq-tblz}
  L_n (\mathbf k) &=
    \frac{2}{m_0}
    \sum_{m \neq n}
      \frac{\textrm{Im}\left[p_{nm}^{(x)} (\mathbf k) p_{mn}^{(y)}(\mathbf k)\right]}{E_{n \mathbf k} - E_{m \mathbf k}},
\\
\label{eq-lz}
  L_n (\mathbf k) &= 
    \frac{2 m_0}{\hbar^2}
    \sum_{m \neq n}
      \textrm{Im}\left[\xi_{nm}^{(x)} (\mathbf k) \xi_{mn}^{(y)}(\mathbf k)\right]
      (E_{n \mathbf k} - E_{m \mathbf k}).
\end{align}

To calculate the contributions of the conduction ($c$) band electron with $\mathbf k_c, s_c$ and the hole ($h$) with $\mathbf k_h, s_h$ to the exciton $g$-factor, we neglect electron-hole Coulomb interactions \cite{Wang2015}. In this case, the exciton Zeeman splitting simplifies to the sum of the Zeeman splittings of the electron and the hole. Using time reversal symmetry which relates the spin and wave vector of the hole to the corresponding spin and wave vector of the empty electron state in the valence ($v$) band ($s_h = -s_v$ and $\mathbf k_h = - \mathbf k_v$), we obtain the exciton $g$-factor as
\begin{equation}
  g^{(cv)} (\mathbf k_c, \mathbf k_v) =
    g_c (\mathbf k_c) - g_v (\mathbf k_v).
\end{equation}

Finally, by reference to the valence band electron with $\mathbf k_v=K$ or $\Gamma$ with spin-up projection $s_v = +1/2$, we discriminate spin-like ($L$) excitons (with $s_c = s_v$) from spin-unlike ($U$) excitons (with $s_c = -s_v$). Their respective exciton $g$-factors are given by:
\begin{align}
  \label{eq-gb}
  g_L^{(cv)} (\mathbf k_c, \mathbf k_v) &=
    2 [ L_c (\mathbf k_c) - L_v (\mathbf k_v)], \\
  \label{eq-gd}
  g_U^{(cv)} (\mathbf k_c, \mathbf k_v) &=
    2 [ L_c (\mathbf k_c) - L_v (\mathbf k_v)] -2g_0.
\end{align}
Using these expressions, we evaluate in the following the exciton $g$-factors from the matrix elements of the orbital angular momentum, $L_c (\mathbf k_c)$ and $L_v (\mathbf k_v)$, obtained within TB and DFT calculations.

\section{TIGHT-BINDING APPROXIMATION}

We begin by calculating the $g$-factors of excitons in both spin-like and spin-unlike configurations of states from different conduction and valence band valleys according to Eq.~\ref{eq-gb} and~\ref{eq-gd} using the results of the eleven-band and six-band TB models of Ref.~\onlinecite{Fang2015} and Ref.~\onlinecite{Rybkovskiy2017}. Without taking into account spin-orbit interactions, we evaluate the matrix elements $L_c (\mathbf k_c)$ and $L_v (\mathbf k_v)$ for the doubly degenerate states of the conduction band at $K$ and $Q$ and the valence band at $K$ and $\Gamma$. By symmetry, the ML energy bands can be classified as odd or even with respect to their reflection in the plane. In the following, we restrict ourselves to even bands with finite contributions to the $g$-factors of excitons with lowest energies \cite{Rybkovskiy2017} and use the notations for orbital indices and hopping integrals from the TB model of \textcite{Fang2015}.

In WSe$_2$ ML, the relevant even bands are formed by even $d$-orbitals of W atoms ($\varphi_6=d_{z^{2}}$, $\varphi_7=d_{xy}$, and $\varphi_8=d_{x^{2}-y^{2}}$) and Se dimers (which include the anti-symmetric combination $\varphi_9=(p_z^A-p_z^B)/\sqrt{2}$ as well as two symmetric combinations $\varphi_{10}=(p_x^A+p_x^B)/\sqrt{2}$ and $\varphi_{11}=(p_y^A+p_y^B)/\sqrt{2}$ of $p$-orbitals of Se). The electron Bloch state within the six-band TB model is approximated as a linear combination of atomic orbitals as \cite{Rybkovskiy2017}:
\begin{equation}
  \psi_{n\mathbf k} (\mathbf r) =
    \sum_{j,l} e^{i\mathbf{k a}_{jl}} c_{nl} (\mathbf k) \varphi_{l} (\mathbf{r - a}_{jl}),
\end{equation}
where $j$ runs over all unit cells, $l = 6 - 11$ enumerates the six even orbitals, and the vectors $\mathbf a_{j,6} = \mathbf a_{j,7} = \mathbf a_{j,8}$ and $\mathbf a_{j,9} = \mathbf a_{j,10} = \mathbf a_{j,11}$ point to the positions of W and Se atoms in the $j$-th unit cell. 

The TB Hamiltonian for the even energy bands is given by a $6\times6$ matrix of the form:
\begin{equation}
  H_{ll'} (\mathbf k) =
    \begin{pmatrix}
      H_\textrm{W-W}  & H_\textrm{W-Se}^\dagger \\
      H_\textrm{W-Se} & H_\textrm{Se-Se}
    \end{pmatrix},
\end{equation}
where the $3\times3$ matrices $H_\textrm{W-W}$, $H_\textrm{Se-Se}$, and $H_\textrm{W-Se}$ describe the interactions between W-W, Se-Se, and Se-W atoms, respectively. 

According to Ref.~\onlinecite{Fang2015}, the six-band TB Hamiltonian has 25 independent fitting parameters: two sets of six hopping integrals $t_{l,l'}^{(1)}$ between the first-neighbor pairs (each for W-W and Se-Se), five parameters $t_{l,l'}^{(5)}$ and four parameters $t_{l,l'}^{(6)}$ for the first- and second-neighbor pairs of atoms of different kinds, and four on-site energies $\epsilon_6$, $\epsilon_{7,8}$, $\epsilon_9$, $\epsilon_{10,11}$. 

With two sets of parameters reproduced from the original work of Ref.~\onlinecite{Fang2015} (TB-1) and Ref.~\onlinecite{Rybkovskiy2017} (TB-2) for ML WSe$_2$ in Tab.~\ref{tab3}, we calculate the matrix elements of the orbital angular momentum using Eq.~\eqref{eq-tblz} and the matrix elements of the momentum operator \cite{Cohen1960,Chang1996}
\begin{equation}\label{eq-tbmom}
  \mathbf p_{nm} (\mathbf k) =
    \frac{m_0}{\hbar}
    \sum_{l,l'}
      c_{nl}^* (\mathbf k) c_{ml'} (\mathbf k)
      \frac{\partial H_{ll'} (\mathbf k)}{\partial \mathbf k}.
\end{equation}

Fig.~\ref{fig5} (a) and (b) show the values of $L_n (\mathbf k)$ within the first Brillouin zone evaluated according to Eq.~\eqref{eq-tblz} for the highest valence band (left panels) and the lowest conduction band (right panels) with the set of parameters from the models TB-1 and TB-2, respectively. The $g$-factors of excitons in specific spin and valley configurations calculated with in Eqs.~\eqref{eq-gb} and \eqref{eq-gd} are listed in Tab.~\ref{tab2}.

\section{Density functional theory}

DFT provides a complementary approach to the calculation of exciton $g$-factors by yielding the energy band structure $E_{n \mathbf k}$ and the matrix elements $\bm{\xi}_{nm} (\mathbf k)$. The angular momenta $L_c (\mathbf k_c)$ and $L_v (\mathbf k_v)$ are obtained from Eq.~\eqref{eq-lz}, and the exciton $g$-factors follow from Eqs.~\eqref{eq-gb} and \eqref{eq-gd}. Our DFT calculations were carried out for ML and BL WSe$_2$ within the generalized gradient approximation (GGA). In brief, the first-principles calculations were performed with the PBEsol exchange-correlation functional~\cite{Csonka-PRB2009} as implemented in the Vienna \emph{ab initio} simulation package (VASP). The van der Waals interactions were considered with the DFT-D3 method with Becke-Johnson damping~\cite{Grimme-JCP2010,Grimme-JCC2011}. The spin-orbit interaction was included at all stages. Elementary cells with a vacuum thickness of $35$~\AA{} were used in order to minimize interactions between periodic images. The atomic positions were relaxed with a cut-off energy of $400$~eV until the change in the total energy was less than $10^{-6}$~eV. The band structure of ML (BL) was calculated on the $\Gamma$-centered $\vec k$ grid of $36 \times 36$ ($18 \times 18$) divisions with $80$ ($160$) bands. 

\begin{figure}[t]   
\centering    
\includegraphics[scale=0.9]{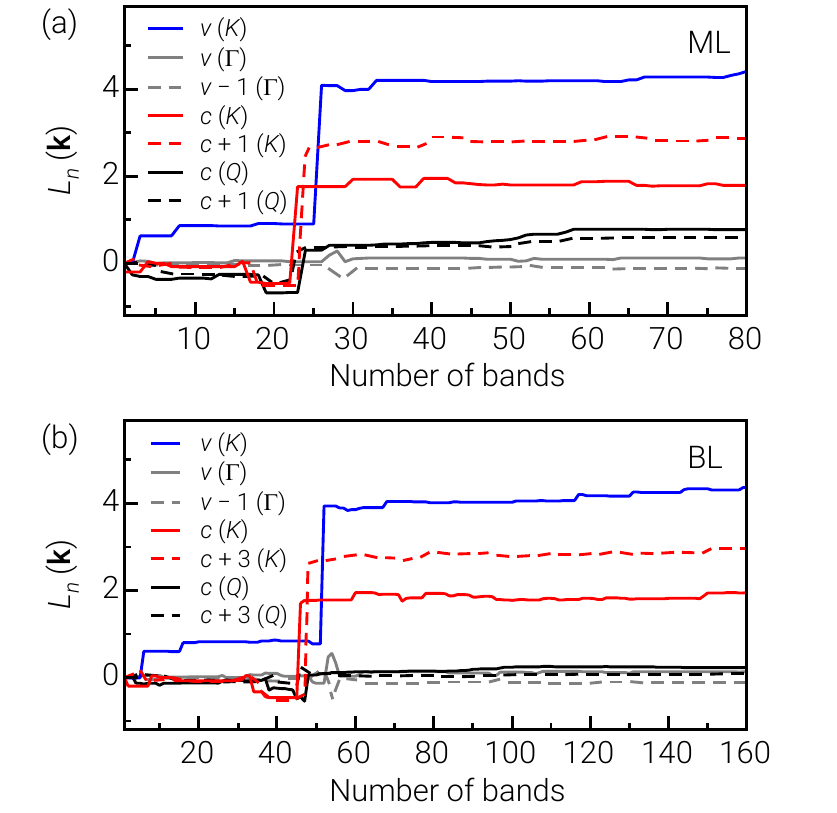}
\caption{Convergence of the electron orbital angular momentum in DFT calculations for the highest valence bands at $K$ and $\Gamma$ and the lowest conduction bands at $K$ and $Q$ in (a) monolayer and (b) bilayer WSe$_2$.}
\label{fig4}
\end{figure}

In Fig.~\ref{fig4}(a) and (b) we show the convergence of the orbital angular momenta $L_n (\mathbf k)$ within our ML and BL calculations as a function of the number of bands taken into account in the sum of Eq.~\eqref{eq-lz}. For the ML, Fig.~\ref{fig4}(a) shows the results for the top-most valance band state $v$ at $K$ (blue solid line), the two highest valence band states $v$ and $v-1$ at $\Gamma$ (grey solid and dashed lines), as well as the two lowest conduction band states $c$ and $c+1$ at $K$ and $Q$ (red and black solid and dashed lines). As the BL bands are doubly degenerate, each $\mathbf k$-point of the Brillouin zone has at least two bands with $L_n (\mathbf k) = L_{n + 1} (\mathbf k)$ or $L_n (\mathbf k) = L_{n - 1} (\mathbf k)$. For the BL in Fig.~\ref{fig4}(b) we consider the same $\mathbf k$-points as for the ML, and show the corresponding bands where the orbital angular momenta have the same sign as in the ML case of Fig.~\ref{fig4}(a).
 
For the orbital angular momenta of these states, convergence is observed above $60$ and $120$ bands in the case of ML and BL in Fig.~\ref{fig4}(a) and (b), respectively, with the factor of two difference related to the doubled number of atoms in BL calculations. We note that the values for the valence band states at $\Gamma$ must vanish by symmetry arguments, whereas our numerical calculations yield $\pm 0.12$ for both ML and BL. This discrepancy is due to a finite number of bands taken into account and can be used to estimate the precision of our numerical calculations. The corresponding bound on the absolute error of the exciton $g$-factors from DFT, given explicitly in Tab.~\ref{tab2} for selected exciton configurations, is thus in the order of $\pm 0.5$. 

\begin{figure}[t!]   
\centering    
\includegraphics[scale=0.9]{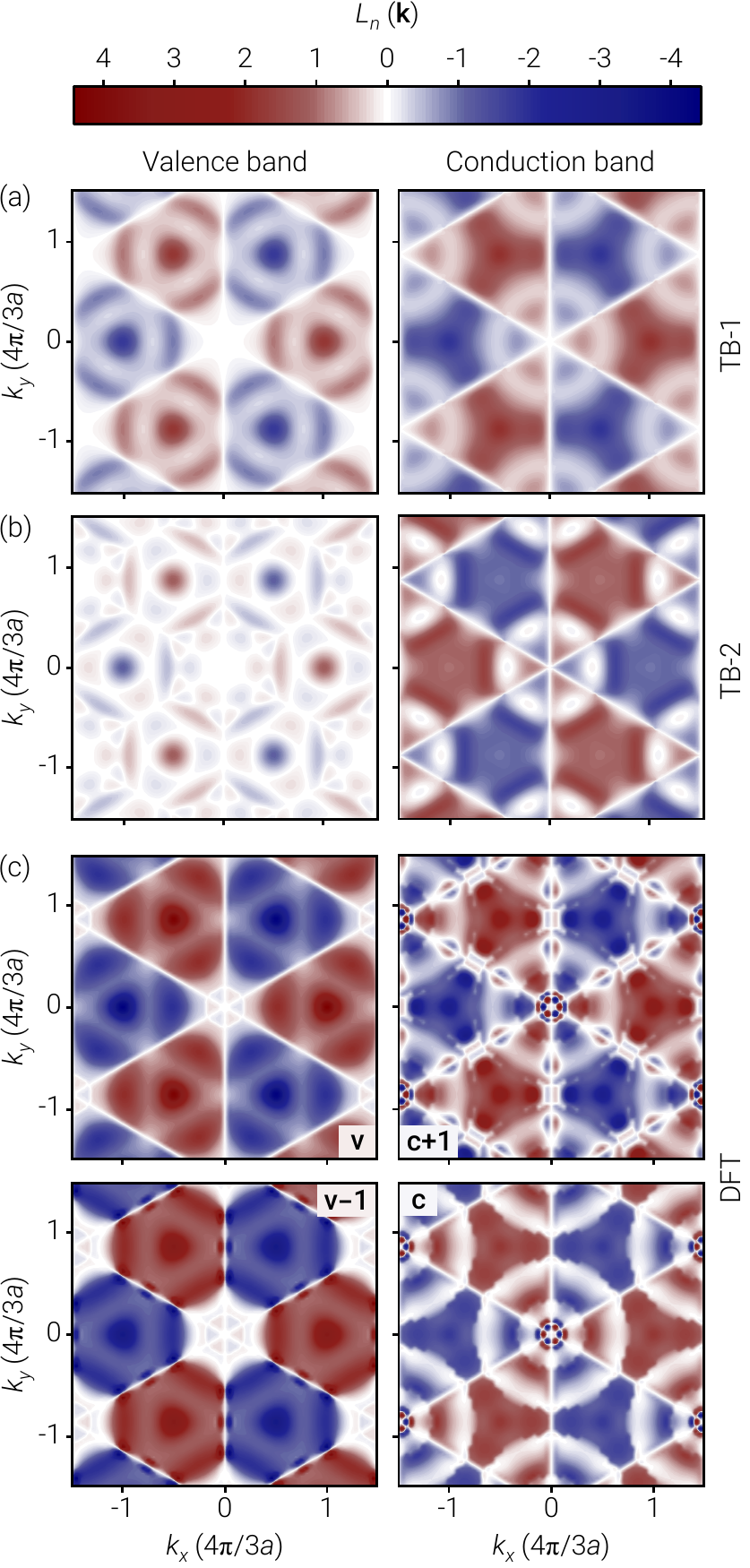}
\caption{Orbital angular momentum in the first Brillouin zone calculated for the top-most valence (left panels) and lowest conduction (right panels) bands from (a) TB-1, (b) TB-2 and (c) DFT. Note that the valence and conduction bands are spin-degenerate in tight-binding without spin-orbit effects, whereas the valence bands $v$ and $v-1$ and the conduction bands $c$ and $c+1$ are split by the respective spin-orbit splitting in DFT calculations.}
\label{fig5}
\end{figure}

As for the TB calculations, we plot in Fig.~\ref{fig5}(c) the DFT results for $L_n (\mathbf k)$ within the first Brillouin zone. Since spin-orbit effects were included at the DFT level, it is instructive to show both spin-orbit split highest valence bands ($v$ and $v-1$) and lowest conduction bands ($c$ and $c+1$). All models agree with respect to the sign, and at least qualitatively, in the overall dependence of $L_v (\mathbf k)$ in the valence bands (left panel). A qualitative agreement between TB-1 and DFT is also apparent for the conduction bands (right panel) in striking contrast to the results of TB-2 that predicts reversed signs for $L_c (\mathbf k)$ throughout almost the entire Brillouin zone. The results of the model also differ substantially from TB-1 and DFT values of $L_v (\mathbf k)$ at the quantitative level. We conclude from this comparison that the correspondence between $L_n (\mathbf k)$ and thus between the exciton $g$-factors obtained from DFT and TB is sensitive to the details of the TB fitting procedures. The main shortcoming of TB models could possibly result from the fact that they are limited to $d$- and $p$-orbitals of W and Se atoms, respectively, and neglect the contributions from other orbitals.

\begin{table}[h!]
\caption{Exciton $g$-factors calculated for ML WSe$_2$ from TB-1, TB-2 and DFT. The $g$-factors in the two right-most columns for BL WSe$_2$ intralayer (Intra) and interlayer (Inter) excitons are both from DFT. Note that without further assumptions the sign of the $g$-factor is only meaningful for zero-momentum spin-like excitons with valley-contrasting dipolar selection rules.}
  \begin{ruledtabular}
  \begin{tabular}{CCC|RRR|RR}
            \text{Exciton}                   & \text{Valley}  & \text{Spin}         & \multicolumn{3}{c|}{ML}                 & \multicolumn{2}{c}{BL}                \\
               & \mathbf{k}_c,\mathbf{k}_v & s_c, s_v & \text{TB-1} & \text{TB-2} & \text{DFT}  & \text{Intra} & \text{Inter} \\
    \hline
    \textit{X$^0$}  & \textit{KK}    &\uparrow\uparrow     &  0.2        & -4.2        & -3.1        & -2.8              & -12.6             \\
    \textit{D$^0$}  & \textit{KK}    &\downarrow\uparrow   & 3.8        & 8.2        & 9.2        & 8.8              & 18.6             \\
    \textit{K$'_{L}$} & \textit{K$'$K} &\uparrow\uparrow     &  7.4        &  0.4        & 12.4        & 12.6              &   2.8             \\
    \textit{K$'_{U}$} & \textit{K$'$K} &\downarrow\uparrow   & 11.4        &  4.4        & 18.5        & 18.6              &   8.8             \\
    \hline
    \textit{Q$_{L}$}  & \textit{QK}    &\uparrow\uparrow     &  3.0        &  2.7        &  7.3        &  8.3              & 8.9       \\
    \textit{Q$_{U}$}  & \textit{QK}    &\downarrow\uparrow   &  7.0        &  6.7        & 11.6  &13.1       &  12.6             \\
    \textit{Q$'_{L}$} & \textit{Q$'$K} &\uparrow\uparrow     &  4.3        &  1.9        & 10.0 & 8.9       &   8.3             \\
    \textit{Q$'_{U}$} & \textit{Q$'$K} &\downarrow\uparrow   &  8.3        &  5.9        & 14.4        & 12.6              & 13.1       \\
    \hline
                        & \textit{K}\,\Gamma  &\uparrow\uparrow     &  3.8        &  1.9        &  5.8        &  5.9              &   3.9             \\
                        & \textit{K}\,\Gamma  &\downarrow\uparrow   &  0.2        &  5.9        &  0.4        &  0.1              &   9.9             \\
                        & \textit{K$'$}\Gamma &\uparrow\uparrow     &  3.8        &  1.9        &  3.6        &  3.9              &   5.9             \\
                        & \textit{K$'$}\Gamma &\downarrow\uparrow   &  7.8        &  2.1        &  9.8        &  9.9              &   0.1             \\
    \hline
                        & \textit{Q}\,\Gamma  &\uparrow\uparrow     &  0.7        &  0.4        &  1.5        &  0.4              &   0.2             \\
                        & \textit{Q}\,\Gamma  &\downarrow\uparrow   &  3.3        &  4.4        &  2.8        &  3.8              &   4.4             \\
                        & \textit{Q$'$}\Gamma &\uparrow\uparrow     &  0.7        &  0.4        &  1.2        &  0.2              &   0.4             \\
                        & \textit{Q$'$}\Gamma &\downarrow\uparrow   &  4.7        &  3.6        &  5.5        &  4.4              &   3.8             \\
  \end{tabular}
  \end{ruledtabular}
  \label{tab2}
\end{table}

With the matrix elements of the orbital angular momenta of the valence and conduction bands in Fig.~\ref{fig5} obtained from TB and DFT for the first Brillouin zone, it is straight forward to calculate the $g$-factors of the lowest-energy ML excitons in various configurations. In Tab.~\ref{tab2} we compare the $g$-factors obtained from TB-1 and TB-2 models with our DFT results for excitons in different configurations of valleys ($\mathbf k_c, \mathbf k_v$) and spins ($\mathbf s_c, \mathbf s_v$, with $\uparrow$ or $\downarrow$ projection along $z$). 

In the top block of Tab.~\ref{tab2}, we list excitons with the hole at $K$ and the electron at $K$ or $K'$ in spin-like and spin-unlike configurations with short exciton notation for zero-momentum bright and dark neutral excitons $X^0$ and $D^0$ and their finite-momentum counterparts $K'_L$ and $K'_U$. The block below shows the results for the spin-like and spin-unlike $Q$-excitons with the electron in $Q$ and the hole in $K$, followed by two blocks without short exciton notation for momentum-indirect excitons composed from electrons in $K$ or $Q$ and holes in $\Gamma$. Note that the sign of the $g$-factor can be determined without further assumptions only for $X^0$ with established dipolar selection rules (the $D^0$ emission is in-plane and linearly polarized). For Zeeman-split momentum-indirect excitons, in contrast, additional symmetry analysis is required to determine their $g$-factor sign from the energetic ordering and polarization of the respective phonon sidebands by taking into account the symmetry of the actual phonons involved in first or higher-order scattering processes that mediate the phonon-assisted PL \cite{HeNatComm2020}. 

\section{Discussion}

First, we discuss the results of our calculations for excitons in ML WSe$_2$. The $g$-factor from TB-2 is very close to the experimental value of $-4$ for $X^0$ \cite{Srivastava2015,Aivazian2015,Wang2015,Mitioglu2015,Stier2018,Koperski2018}, whereas our DFT model predicts $-3.1$, and the result from TB-1 of $0.2$ is completely off. We note that the disagreement between experiment and DFT is actually surprisingly small given the sample-to-sample variations in experimental reports \cite{Srivastava2015,Aivazian2015,Wang2015,Mitioglu2015,Stier2018,Koperski2018} and the limited number of bands included in our DFT calculations. We expect the agreement to improve with the number of bands and approach the excellent agreement in the $g$-factor of spin-dark excitons with $g \simeq 9.4$ in experiment \cite{RobertPRB2017} and $9.2$ in DFT. 

The states $K'_L$ and $K'_U$, which are the momentum-indirect counterparts of $X^0$ and $D^0$, respectively, exhibit different $g$-factors with large values of $12.4$ and $18.5$ not predicted by either of the two TB models. We find a similar discrepancy between large ($7.3$ to $14.4$) and small ($1.9$ to $8.3$) $g$-factor values from DFT and TB for $Q$-momentum excitons, whereas all theories agree on the smallness of $g$-factors for excitons with the hole at $\Gamma$. As expected, the $g$-factors of intralayer excitons in BL WSe$_2$ are close to the values of the corresponding ML excitons \cite{Arora2018}. In addition to intralayer excitons, the BL hosts interlayer counterparts (\textit{e.g.} intralayer $Q_{L}$ and interlayer $Q'_{L}$, intralayer $Q_{U}$ and interlayer $Q'_{U}$, an so on) that exhibit the same $g$-factors in our model which neglects Coulomb corrections for intralayer and interlayer excitons. 

By providing explicit $g$-factor values for momentum-indirect excitons, our DFT results complement the experimental observations in ML and BL WSe$_2$. In the framework of neutral MLs, however, they do not resolve the ambiguity between the two competing explanations of the peak $M^0_1$. The assignment of the peak as a phonon sideband of $Q$-momentum excitons \cite{Lindlau2017ML}, on the one hand, is consistent with the $g$-factors of $7.3$ and $14.4$ for $Q_L$ and $Q'_U$ states in Tab.~\ref{tab2} (note that $Q_U$ and $Q'_L$ excitons, $250$~meV above degenerate $Q_L$ and $Q'_U$ states \cite{Deilmann2019}, are irrelevant in this context) and the structured peak $M^0_1$ in Fig.~\ref{fig2} with a $g$-factor of $11.5$. On the other hand, the interpretation of the peak as direct PL emission by momentum-dark $K'_L$ excitons \cite{HeNatComm2020} is also consistent with the theoretical $g$-factor of $12.4$ from DFT. Our DFT results also identifies $K\Gamma$ and $Q\Gamma$ with small $g$-factors as potential candidates to explain the bright PL peak between $M^-_1$ and $D^-$ in the negatively charged regime of high-quality samples with narrow spectra \cite{LiuPRR2019,HeNatComm2020}. 

For the neutral BL, our results help to rule out $Q\Gamma$ excitons and suggest spin-unlike interlayer $QK$ and intralayer $Q'K$ exciton reservoirs rather than $K'\Gamma$ as a joint origin of phonon sidebands $B^0_1$, $B^0_2$ and $B^0_3$ \cite{Lindlau2017BL}. Whereas a detailed assignment of the neutral BL peaks to the specific reservoirs and phonon sidebands is yet to be developed, the values of the exciton $g$-factors in the charged regime can be understood, as in the ML case, by regarding the additional electron in the charged complex simply as a spectator to the Zeeman effect of the neutral finite-momentum exciton reservoir. 

\section{Summary and conclusions}

In summary, our work provides exciton $g$-factors for neutral and charged ML and BL WSe$_2$ from both experiment and theory. For neutral and charged ML WSe$_2$ it complements previous experimental findings by theoretical calculations of the $g$-factors for momentum-indirect excitons in different configurations of spins and valleys. We find overall very good quantitative agreement with experiment for theoretical $g$-factor values obtained from first-principles calculations, whereas TB methods fail at matching the experimental values for some of the versatile exciton species in MLs. For BL WSe$_2$, our work adds new insight into the origin of PL peaks on the basis of theoretical $g$-factor values. In the broad context of research on layered semiconductors and their applications, the theoretical aspects of our work provide new guidelines for magneto-optical studies of single-layer TMDs, homo- or hetero-bilayer systems, and other realizations of TMD-based van der Waals heterostructures.

\begin{acknowledgments}
The authors thank M.~M.~Glazov, T.~Deilmann and P.~Hawrylak for fruitful discussions. This research was funded by the European Research Council (ERC) under the Grant Agreement No.~772195, the Volkswagen Foundation, and the Deutsche Forschungsgemeinschaft (DFG, German Research Foundation) under Germany's Excellence Strategy EXC-2111-390814868. S.~Yu.~K. acknowledges support from the Austrian Science Fund (FWF) within the Lise Meitner Project No. M 2198-N30. A.~S.~B. has received funding from the European Union's Framework Programme for Research and Innovation Horizon 2020 (2014--2020) under the Marie Sk{\l}odowska-Curie Grant Agreement No.~754388, and from LMU Munich's Institutional Strategy LMUexcellent within the framework of the German Excellence Initiative (No.~ZUK22). A.~H. acknowledges support from the Center for NanoScience (CeNS) and the LMUinnovativ project Functional Nanosystems (FuNS). Growth of hexagonal boron nitride crystals was supported by the
MEXT Element Strategy Initiative to Form Core Research Center, Grant Number JPMXP0112101001 and the CREST(JPMJCR15F3), JST.
\end{acknowledgments}


\bibliography{bibliography_g-factor_v4}

\end{document}